\documentclass[aps,pre,twocolumn,amsmath,amssymb]{revtex4-1}
\usepackage{graphicx}
\usepackage{epsfig}

\begin{document}

\title{Pattern in non-linearly coupled network of identical Thomas oscillators}



\author{Vinesh Vijayan}
\email{vineshvijayan@nitrkl.ac.in}
 \affiliation{Department of Physics \& Astronomy, National Institute of Technology Rourkela, India}
\author{B. Ganguli}%
 \email{biplabg@nitrkl.ac.in}
\affiliation{Department of Physics \& Astronomy, National Institute of Technology Rourkela, India}


\date{\today}

\begin{abstract}
We have investigated synchronized pattern in a network of Thomas oscillators coupled with sinusoidal nonlinear coupling. Pattern like chimera states are not only observed for many non-locally coupled oscillators but there is a signature of it even for locally coupled few oscillators. For certain range of intermediate coupling, clusters are also observed. These patterns do resemble with motion of real self propelled coupled dynamical systems. 
\end{abstract}
\maketitle

\section{Introduction}
Complexity frontier deals with the study of collective phenomena where one is interested in understanding overwhelmingly intriguing emergent behaviors which comes out as a fact of complex interactions of large number of interacting parts. Spatio-temporal pattern formation which is the emergence of patterns of large scale behavior is one such fascinating phenomena in complexity science. Emergent phenomena of this sort sheds light on to a wide range of collective phenomena and also to its better understanding. One such phenomena of recent interest is that of chimera states in an ensemble of coupled oscillators. It is the coexistence of coherent and incoherent patterns emerging out on otherwise identical settings of oscillators and coupling topology. This new dynamic state emerges out because of a spontaneous symmetry breaking.\\
Signatures of spatio-temporal evolution in a system of coupled nonlinear oscillators were found and was named domain like spatial structure  much earlier than Kuramato and co-workers\cite{Umberger}. Until Kuramato there was no theoretical explanation for this phenomena  and theoretical exploration by Strogartz, who coined the name chimera state, paved the way for further understanding \cite{Kuramato,Abraham}. There after a plethora of studies followed on chimera states starting from the investigation on spontaneous symmetry breaking \cite{Adilson,Mark}. Studies were extended to much more general situations which included time delay and phase lag \cite{Sethia,Nan}. It was observed that these dynamical states can be observed in non-locally coupled chaotic maps, Rossler oscillators, neural system and coupled complex Stuart Landau equations \cite{Omel,Omel1,Sagu,Gopal,Oleh,Oleh1}. The role of coupling was further investigated from nonlocal coupling to global coupling  and also with some subpopulation model  as well as with some randomized coupling  to unveil the effect of coupling topology \cite{Oleh,Abraham1,Nan1}. Recent theoretical developments on chimera states included the study of directed flow of information within and between coupled subpopulations of phase oscillator network in the chimera state \cite{Nicolas}. The observation of amplitude chimera states against the conventional phase chimera  is also a notable advance in the field preceded by the identification of imperfect amplitude chimera  due to attractive and repulsive interactions in a non-locally coupled Stuart Landau oscillators \cite{Tanmay,Sathya}. At the frontier one can see the extension of network topology to multilayer and multiplex networks for understanding chimera states   \cite{Aleksei,Jakub,Iryana}. Practical application of chimera states are demonstrated in the laboratory in the last decade apart from the theoretical and numerical analysis \cite{Lennart,Carlo,Yun}.\\
Observations of chimera states are  achieved in network of non-linear oscillators mostly with linear coupling scheme. Kuramoto model being the exceptional, where nonlinear sinusoidal coupling itself provide non-linearity to the system of linear phase oscillators. Non-linearity in the coupling scheme can not be ruled out in real systems described by chaotic dynamics.\\
Rene Thomas introduced a particularly simple 3D flow with two important properties. The system of equations are cyclic under the interchange $x,y,z$ coordinates and there is only one dissipation parameter  $b$. The model is given by
\begin{eqnarray}
\frac{dx}{dt} = -bx +siny\\
\frac{dy}{dt} = -by +sinz\\
\frac{dz}{dt} = -bz +sinx 
\end{eqnarray}
It was originally proposed for modeling feedback circuit\cite{Thomas}. Subsequently mathematical modeling based on feed back circuit is also found to be useful for many biological systems like cell differentiation\cite{thomas-kauf2} and regulatory network\cite{thomas-kauf1}. Therefore Thomas system in general is a suitable model to understand many biological systems\cite{Vasi} and collective motion of self propelled active agents. Present understanding of collective motion of active agents comes from the work on modeling of self driven active systems\cite{ T_vicsek} and Active Brownian particles\cite{cates}. In such a model there is a tendency to align velocities with their neighboring counterpart. Such alignment of velocities gives rise to synchronized pattern in the collective motion of self propelled agents. Therefore coupled Thomas system may be a suitable model for understanding of collective dynamics of Active Brownian particles which have velocity-velocity correlation. The variables, $x, y, \& z$ in this model then represent components of velocity. Motion of an active Brownian particle is considered as random motion.  In a recent work collective motion of active Brownian particles is modeled with stochastic dynamics with non-linear phase coupling\cite{Ait18}. Therefore Thomas system in its chaotic regime with non-linear coupling will be suitable for modeling real system like active Brownian particles.\\
The system exhibits rich dynamic phenomena for different values of $b$.  It has stable complex quasi-periodic oscillation, but closed to chaotic dynamics for $b=0.1998$. The system has a single chaotic attractor for $b=0.18$ and chaotic behavior without any attractor(Labyrinth chaos, chaotic walk) in the case of $b=0.0$. In the latter case the system becomes conservative with infinite lattice of unstable fixed points. Physically the model stands for a particle moving in a force field with frictional damping under the action of some external source of energy\cite{Thomas,Sprott,Rowlands,Vasi}, which makes the particle an active agent.\\

In a recent work\cite{vin}, we had investigated synchronized states of two coupled Thomas oscillators and compared our result for linearly and non-linearly coupled system. Non-linearly coupled system not only shows complete synchronization like observed in linearly coupled, it also shows windows of lag and anti-lag synchronization. Such result is indeed  plausible in real systems as indicated in the work of \cite{Ait18}. This result shows that non-linearly coupled Thomas system may show interesting synchronized pattern in a network and useful to understand collective dynamics in real systems.\\

This paper attempts an investigation of possible generation of synchronized pattern in a network of Thomas oscillators in a ring topology. We are particularly interested to investigate the effect of sinosoidal non-linear coupling scheme on the nature of coupling topology, i.e., global, local and non-local couplings. We are also looking at the effect of different dynamical behavior of individual  unit on pattern formation by considering different values of dissipation parameter $b$. Finally we show the importance of our result in the context of active Brownian particles motion in a real system. The definite quantitative charaterization of patterns is based on the recently developed statistical tools by Gopal et al\cite{Gopal}.

\section{Modeling And Observation of Cluster States}

Non-locally coupled Thomas oscillators with nonlinear coupling function can be modeled using the following equation,
\begin{eqnarray}
{\bf \dot{x_i}} = {\bf f_i(x_i)} + \frac{{\bar{\epsilon}}}{2P} \sum_{j=i-P}^{j=i+P} sin( x_j -x_i)
\end{eqnarray}
where we consider ${\bf f_i(x_i) }\hspace{0.1cm} \varepsilon \hspace{0.1cm}\Re^{3N}$ as given by (1) with ${\bf x_i}=(x_i,y_i,z_i),\hspace{0.1cm} {\bf \dot{x_i}} =(\dot{x_i},\dot{y_i},\dot{z_i})\hspace{0.1cm}\varepsilon \hspace{0.1cm}\Re^{3N}$. The index ${\bf i}$ stands for labeling a particular oscillator(${\bf i}=1,2,3.....n$). ${\bf \bar{\epsilon}}$ is the coupling matrix given by
\begin{equation}
{\bf \bar{\epsilon}} = \epsilon \begin{bmatrix}
1 & 0 & 0\\
0 & 0 & 0\\
0 & 0 & 0\\
\end{bmatrix}
\label{this}
\end{equation} 
where $\epsilon$ is the coupling strength. N is the number of oscillators and P is the number of nearest neighbors on both sides of the ${\bf i}^{th}$ oscillator. The coupling radius is defined as $r =P/N$.\\
Characterization of synchronized pattern is based on the work of Gopal et al\cite{Gopal}. For this purpose we calculate standard deviation of relative differences of state variables between neighboring oscillators which are defined  as: ${\bf w_{j,i}} = (w_{x,i},w_{y,i},w_{z,i}) = (x_i-x_{i+1},y_i-y_{i+1},z_i - z_{i+1}) \hspace{0.1cm}\varepsilon \hspace{0.1cm}\Re^{3N}$. When the $i^{th}$ and the $ i+1^{th}$ oscillators are oscillating coherently then the value of  ${\bf w_{j,i}}$ will be minimum. In the case of incoherent oscillations ${\bf w_{j,i}}$ gets distributed between the upper and lower bounds of the allowed values of the state variables. For the chimera states some of the  ${\bf w_{j,i}}$ may have the same value but the others will get distributed in the above mentioned bounds. The quantification of the synchronized state is defined by the standard deviation for the asymptotic states as
\begin{eqnarray}
 \sigma_j = \Biggr \langle \sqrt{\frac{1}{N} \sum_{i=1}^N [{ w_{j,i}} -\langle { w_j} \rangle ]^2}\hspace{0.1 cm} \Biggr \rangle _t
\end{eqnarray}   

where $<...>_t$ denotes  time average and $ j= x,y,z $. $\langle w_j \rangle = \frac{1}{N} \sum_{i=1}^N w_{j,i}(t)$. $ \sigma_j $ will take zero values for coherent states but non zero values for both incoherent and chimera states. It turns out that the standard deviation for asymptotic states is not a good quantitative measure to distinguish between chimera states and incoherent sates. In order to overcome this difficulty we divide the oscillators in to n(even) number of bins of equal length $M=\frac{N}{n}$ and define a local standard deviation, $\sigma_j(m)$ as follows
\begin{eqnarray}
 \sigma_j(m) = \Biggr \langle \sqrt{\frac{1}{M} \sum_{j=n(m-1)+1}^{mM} [{ w_{j,i}} -\langle { w_j} \rangle ]^2}\hspace{0.1 cm} \Biggr \rangle _t
\end{eqnarray} 
where $m=1,2,....n$. The strength of incoherence is then defined by  
\begin{eqnarray}
 {\bf SI } = 1 - \frac{\sum_{m=1}^n s_m}{n}
\end{eqnarray} 
where $s_m$ is the Heaviside step function, i.e., $s_m =\Theta(\delta -\sigma_j(m)) $. $\delta $ is a predefined threshold and is reasonably small. It is suitably chosen as a small percentage of the difference between the largest and the lowest values of state variable. $s_m$ is so defined that ${\bf SI}$ is one for incoherent state while zero for coherent state. For chimera state the value of ${\bf SI}$ will be between zero and one. To distinguish further between chimera and multi chimera state one more measure is defined from the distribution of $s_m$ as follows
\begin{eqnarray}
 {\bf DM } = \frac{\sum_{i=1}^n{|s_i - s_{i+1}|}}{2}\hspace{0.2cm}, s_{n+1} =s_1
\end{eqnarray}
and is known as the discontinuity measure.The value of ${\bf DM}$ is one for single chimera but will be an integer greater than one for multi- chimera state.
\section{Global and non-local coupling schemes}
\subsection{Complex Quasi periodic oscillations}
\begin{figure}[h]
\begin{center}
\includegraphics[width = 3.5 in]{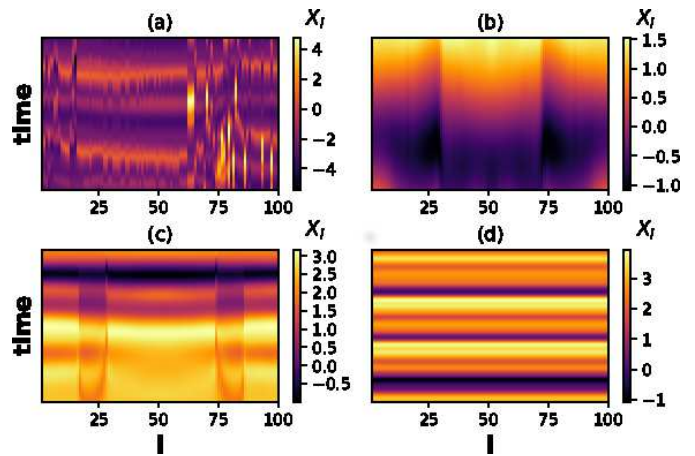}
\end{center}
\caption{Space time plots  of non locally coupled Thomas system($b = 0.998$) for different values of coupling coefficient in terms of state variable $X_i$: (a) incoherent state ($\epsilon = 0.8$),(b) chimera state ($\epsilon = 1.8$),(c) multi chimera state ($\epsilon = 2.7$), (d) coherent state ($\epsilon = 3.1$). The coupling radius is fixed at $r = 0.2$ and $N = 100$.}
\label{AB}
\end{figure}

\begin{figure}[h]
\begin{center}
\includegraphics[width = 3 in]{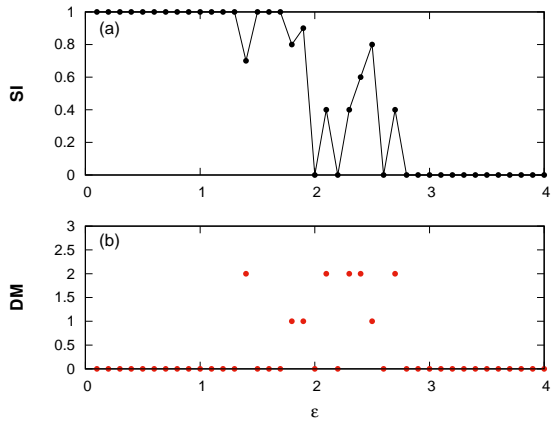}
\end{center}
\caption{(a) Strength of incoherence ({\bf SI}) and (b) discontinuity measure ({\bf DM}) versus coupling strength $\epsilon$ for non locally coupled Thomas system for $b = 0.1998, N = 100$ and $r = 0.2$.}
\label{AC}
\end{figure}

We first consider the case of complex quasi periodic oscillations of Thomas oscillator which happens to be for the value of damping coefficient $b = 0.1998$. Number of oscillators is taken to be $100$. Coupling radius $r$ is then equal to $0.5$ for global coupling and $0.01$ for only first nearest neighbor local coupling. The network quickly achieve complete synchronized state for very low value of coupling strength for global coupling as expected. Similar result is also obtained when coupling scheme is linear and  diffusive.\\ For the case of non-local coupling, we consider a reasonable intermediate value of $r$($=0.2$). The coupled system remains incoherent up to a lower bound of coupling value,  $\epsilon_i = 1.3$. From here onwards within a range of $\epsilon = 1.3$ to $2.8$, the system undergoes transition through cluster, chimera and multi-chimera states. After reaching chimera states at $\epsilon = 1.4$ the system becomes incoherent for $\epsilon = 1.5 - 1.7$. We have multi chimera states for $\epsilon = 1.4 , 2.1 ,2.3 , 2.4 , 2.7$ ( $DM=2$ ) and chimera states for $\epsilon = 1.8 , 1.9 , 2.5$ ($DM=1$). The strength of incoherence and discontinuity measures are shown in  Figure(\ref{AC}). The system finally settles down to coherent state(in this case it is a complete synchronization) from an upper bound of $\epsilon_c = 2.8$.
\begin{figure}[h]
\begin{center}
\includegraphics[width = 3.5 in]{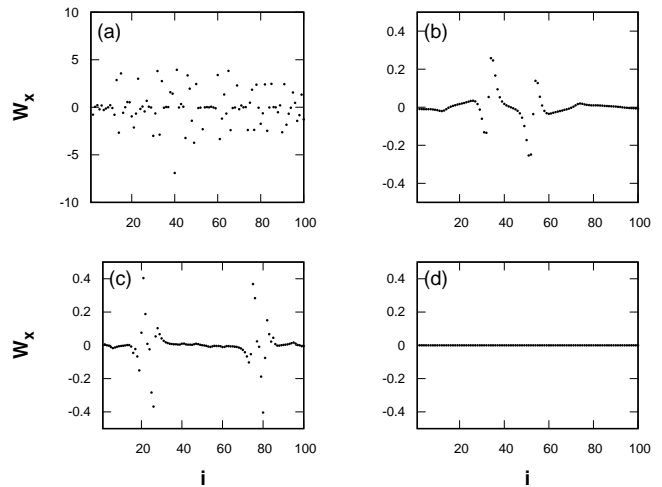}
\end{center}
\caption{Snapshots of non locally coupled Thomas system($b = 0.998$) for different values of coupling coefficient in term of new state variable ${\bf W_x} = W_{x,i}$: (a) incoherent state ($\epsilon = 0.8$),(b) chimera state ($\epsilon = 1.8$),(c) multi chimera state ($\epsilon = 2.7$), (d) coherent state ($\epsilon = 3.1$). The coupling radius is fixed at $r = 0.2$ and $N = 100$.}
\label{AA}
\end{figure}
Figure(\ref{AA}) shows the snapshots of incoherent, chimera, multi chimera and coherent states for $ \epsilon = 0.8,1.8,2.7 $ and $3.1$ respectively. For $\epsilon = 1.8$, there is clearly a cluster from $i = 30$ to $i = 70$, where the new state variable gets distributed showing the emergence of chimera states. Similarly for $\epsilon = 2.7$, we have two clusters centered around $20^{th}$ and $80^{th}$ oscillators. Therefore in the case of complex quasi periodic oscillations, the transition from  incoherent to coherent(complete synchronized) state is through cluster, chimera and multi chimera states. Spatio-temporal patterns are shown in Figure(\ref{AB}) for this case.

\begin{figure}[h]
\begin{center}
\includegraphics[width = 3.5 in]{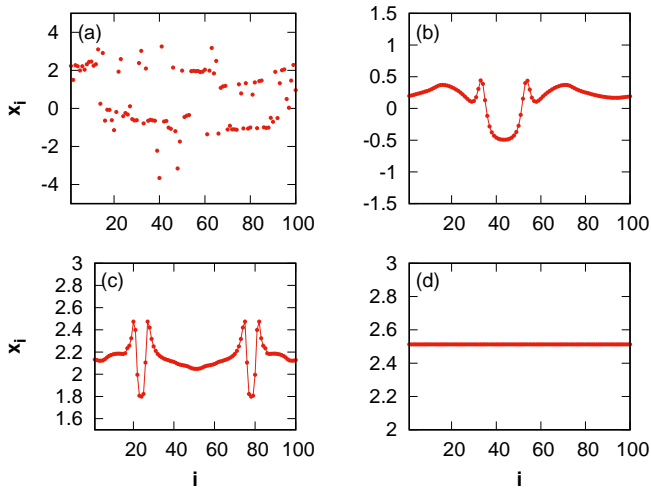}
\end{center}
\caption{Snapshots of non locally coupled Thomas system($b = 0.998$) for different values of coupling coefficient in term as of state variable ${\bf x_i}$: (a) incoherent state ($\epsilon = 0.8$),(b) chimera state(single crater) ($\epsilon = 1.8$),(c) multi chimera state(double craters) ($\epsilon = 2.7$), (d) coherent state ($\epsilon = 3.1$). The coupling radius is fixed at $r = 0.2$ and $N = 100$.}
\label{AD}
\end{figure}

 The snapshot of state variable distribution with respect to oscillators(Figure(\ref{AD})) shows crater shaped distributions. These crater shapes correspond to chimera states shown in Figure(\ref{AA}). A large cluster of about 70 oscillators is formed at $\epsilon=1.8$ (single chimera case) first. On increasing  $\epsilon$, single cluster gets destabilized before two clusters are formed (multi chimera case) at $\epsilon = 2.7$.  

\subsection{Stable Chaotic oscillations} 
\begin{figure}[h]
\begin{center}
\includegraphics[width = 3.5 in]{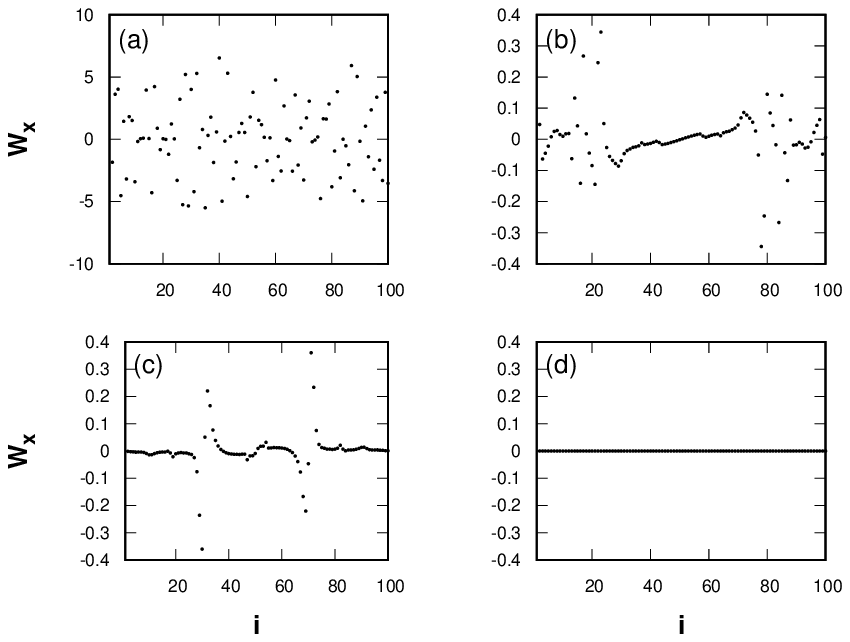}
\end{center}
\caption{Snapshots of non locally coupled Thomas system($b = 0.18$) for different values of coupling coefficient in terms of new state variable ${\bf W_x} = W_{x,i}$: (a) incoherent state ($\epsilon = 0.6$),(b) chimera state ($\epsilon = 2.5$),(c) multi chimera state ($\epsilon = 2.8$), (d) coherent state ($\epsilon = 3.6$). The coupling radius is fixed at $r = 0.2$ and $N = 100$.}
\label{AE}
\end{figure}

\begin{figure}[h]
\begin{center}
\includegraphics[width = 3.5 in]{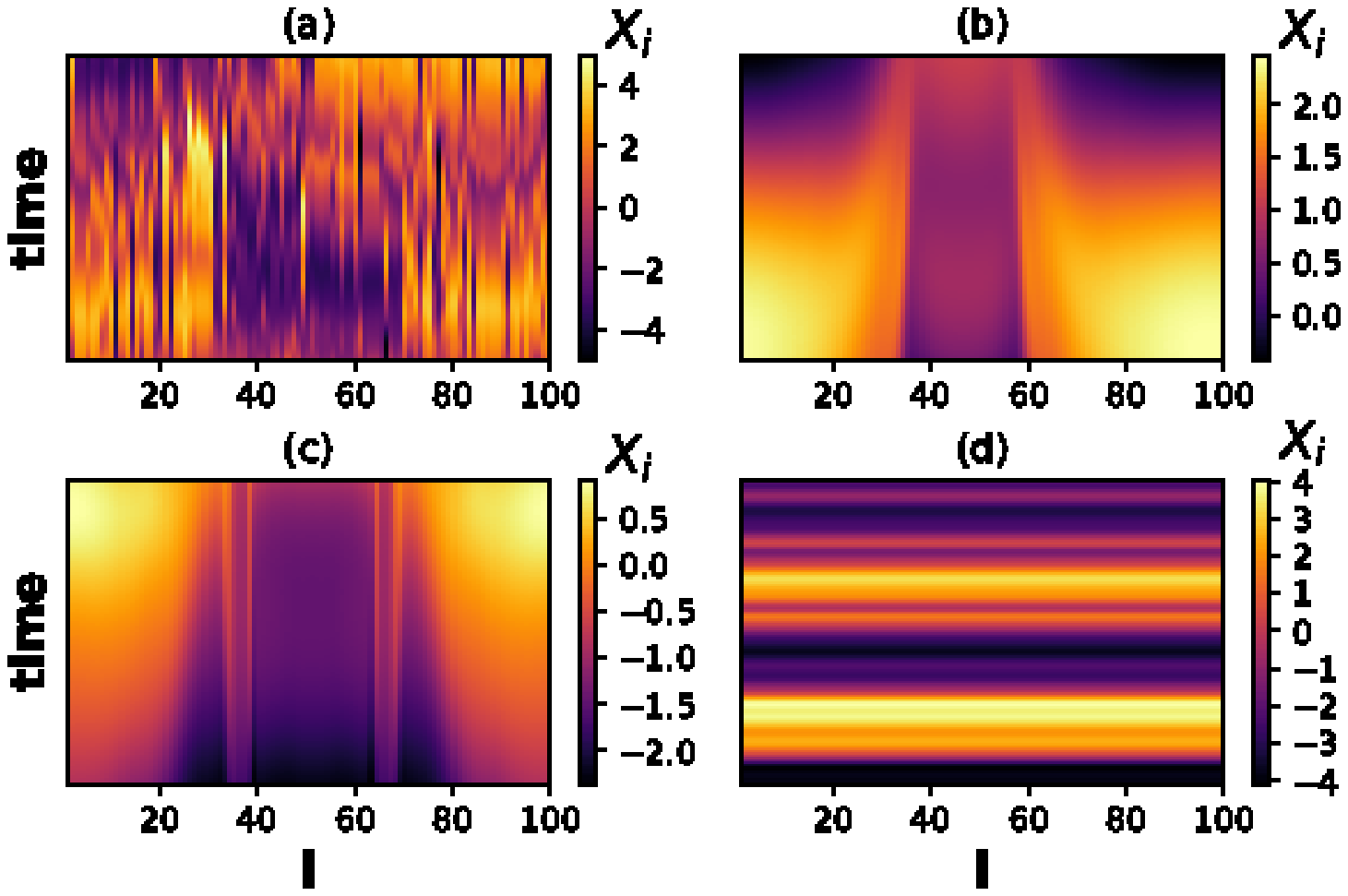}
\end{center}
\caption{Space time plots of non locally coupled Thomas system($b = 0.18$) for different values of coupling coefficient in terms of state variable $X_i$: (a) incoherent state ($\epsilon = 0.8$),(b) chimera state ($\epsilon = 1.8$),(c) multi chimera state ($\epsilon = 2.7$), (d) coherent state ($\epsilon = 3.1$). The coupling radius is fixed at $r = 0.2$ and $N = 100$.}
\label{AF}
\end{figure}

\begin{figure}[h]
\begin{center}
\includegraphics[width = 3 in]{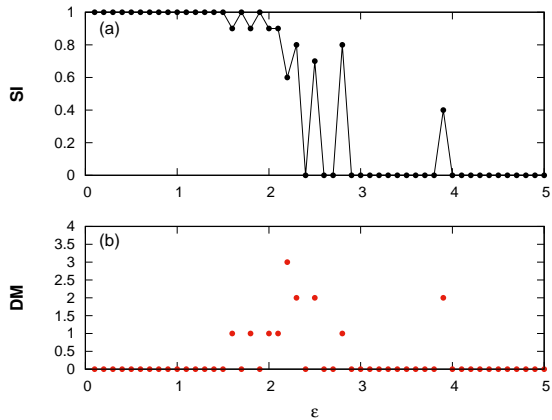}
\end{center}
\caption{(a) Strength of incoherence ({\bf SI}) and (b) discontinuity measure ({\bf DM}) versus coupling strength $\epsilon$ for non locally coupled Thomas system for $b = 0.18, N = 100$ and $r = 0.2$.}
\label{AG}
\end{figure}

The stable chaotic oscillations of the uncoupled Thomas oscillator happens to be at $b=0.18$. In this case also, the system quickly achieve complete synchronization for a very weak value of coupling strength when they are globally coupled.\\
In the case of non local coupling with radius, fixed at $0.2$, the coupled system undergoes incoherent to coherent transition through cluster, chimera, multi  chimera \& complete coherent states. The system remains incoherent up to the lower bound of $\epsilon_i = 1.5$. On further increasing $\epsilon$, the system goes to the transition regime where it undergoes cluster, chimera, multi chimera \& and a complete coherent cluster states until and upper bound of $\epsilon_c =4.0$. From $\epsilon_c = 4.0$ onwards the system attains stable coherence where oscillators synchronize completely. Therefore unlike in the case of complex quasi periodic case, the transition range in $\epsilon$ has a small part ($\epsilon = 2.9$ to $3.9$) of complete coherence and the transition range is also got widened on decreasing $b$. Figure(\ref{AG}) shows multi chimera states for $\epsilon = 2.2, 2.3, 2.5$ \& $3.9$ ($DM=2$) and chimera states for $ \epsilon = 1.6, 1.8, 2.0, 2.1$ \& $2.8$.\\
 Figure(\ref{AE}) shows the snap shots of incoherence, chimera, multi chimera  and coherent states  at $ \epsilon = 0.6, 2.5, 2.8$ \& $3.6$ respectively. There is a cluster of oscillators between $35$ to $60$ which behaves coherently and the remaining oscillators remain incoherent. Two clusters are formed, one centered around $30^{th}$ oscillator and the other around the $70^{th}$ oscillator for $\epsilon = 2.8$. The size of clusters is less compared to quasi-periodic case.
The spatio-temporal pattern in this case is shown in the Figure(\ref{AF}).\\ 
In this case also the state variable distribution with respect to oscillators shows cracker shaped distribution at position where chimera appears.
\subsection{Labyrinth Chaotic oscillations}
\begin{figure}[h]
\begin{center}
\includegraphics[width = 3.5 in]{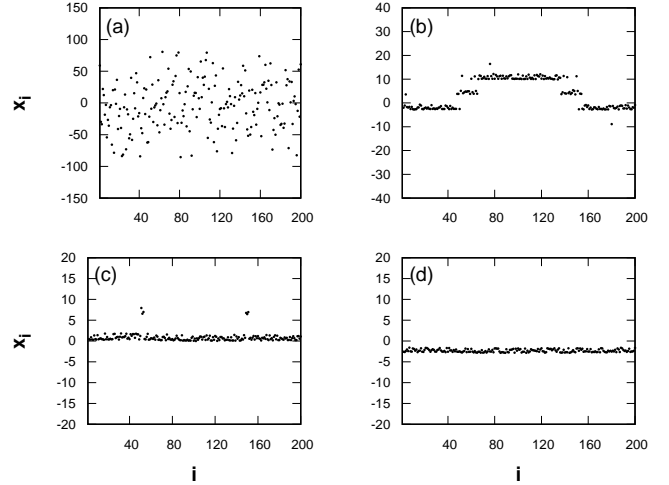}
\end{center}
\caption{Snapshots of non locally coupled Thomas system($b = 0.0$) for different values of coupling coefficient in terms of  state variable ${\bf x_i}$: (a) incoherent state ($\epsilon = 0.7$),(b) cluster state ($\epsilon = 1.4$),(c) cluster state ($\epsilon = 1.6$), (d) coherent state ($\epsilon = 2.0$). The coupling radius is fixed at $r = 0.2$ and $N = 200$.}
\label{AH}
\end{figure}

\begin{figure}[h]
\begin{center}
\includegraphics[width = 3 in]{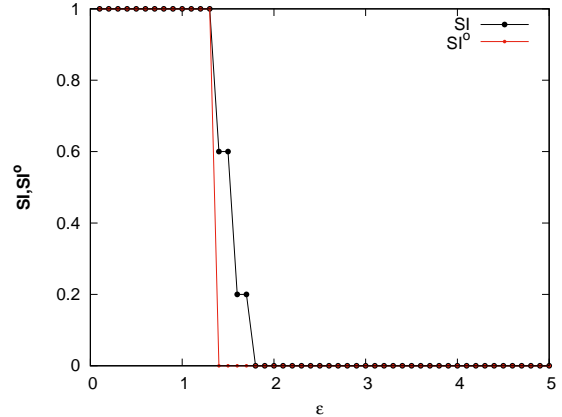}
\end{center}
\caption{(a) Strength of incoherence ({\bf $SI,SI^0$}) with and without removing discontinuity for non locally coupled Thomas system for $b = 0.0, N = 200$ and $r = 0.2$.}
\label{AI}
\end{figure}

For $b = 0$ the uncoupled Thomas system shows conservative chaotic oscillations without an attractor. The motion of the trajectory in phase space reassembles like random walk, but in a purely deterministic manner. When such oscillators are coupled non locally the system undergoes incoherence to coherent transition through cluster states only. Therefore there is no chimera states in this case. Since no chimera states are observed while taking $N = 100$ and clusters are also not very clear, we repeated our calculation for higher number of $N(= 200)$. With $N = 200$, the coupled system remains incoherent upto the lower bound of  $\epsilon_i = 1.3$. The system attains imperfect coherence at $\epsilon_c = 1.8$. In the range of $\epsilon = 1.4$ to $1.7$, the coupled system undergoes transition through imperfect cluster states as it is clear from the Figure(\ref{AH}), where state variable $x_i$ is plotted instead of the difference $W_x$. These imperfect clusters are clearly  visible due to the finite discontinuities in the ${\bf x_i}$ at some specific locations. To confirm the absence of chimera states, we calculated $SI$ using the method of removable singularity\cite{Mal}. Using this method it is possible to distinguish between the cluster states and chimera states. Just like chimera states the cluster states also show a value between zero and one in the $SI$ plot. The Figure(\ref{AI}) shows $SI$ vs $\epsilon$ plot before and after removing the discontinuity.  $SI$ and $SI^0$(after removing discontinuity) are equal for coherent, incoherent, chimera, multi chimera states whereas for  cluster states $SI^0$ is zero but $SI$  assumes nonzero value. Therefore the Figure(\ref{AI}) shows there are only clusters in transition region. The transition from incoherent to coherent state resembles like first order phase transition.

 The transition range is very small compared to the previous two cases. One very important difference between this case and the previous two is that pattern is similar in all the variables in the previous two cases where as no pattern is observed in $y$ and $z$ components in the present case. This may due to the complete absence of diffusive term in $y$ and $z$ component whereas diffusion comes in $x$ through coupling. 

\section{Nearly locally coupled Thomas Oscillators}
\begin{figure}[h]
\begin{center}
\includegraphics[width = 3.5 in]{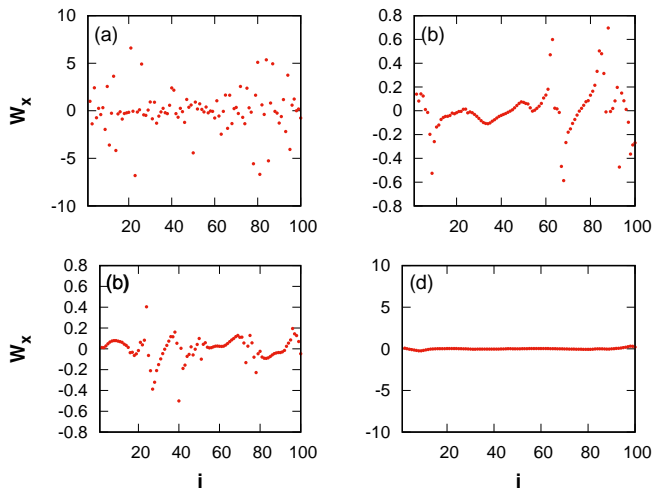}
\end{center}
\caption{Snapshots of nearly locally coupled Thomas system for different values of coupling coefficient in terms of new state variable ${\bf W_x} = W_{x,i}$: (a) incoherent state ($\epsilon = 0.8$),(b) chimera state ($\epsilon = 2.0$),(c) multi chimera state ($\epsilon = 2.2$), (d) coherent state ($\epsilon = 2.6$). The coupling radius is fixed at $r = 0.03$ and $N = 100$.}
\label{AJ}
\end{figure}

When the coupling radius is $r < 0.03$, there is no coherent pattern emerge at all for both linear and non-linear coupling and for all types of dynamics when coupling strength is low or moderate($\epsilon = 20.0$). For the case of $b = 0.1998$ with $r = 0.03$, coherent pattern starts emerging when coupling is sinusoidal non linear. No such pattern emerge when coupling is linear for this value of $r$. The Figure(\ref{AJ}) shows there is chimera state for $\epsilon = 2.0$, multi-chimera for $\epsilon = 2.2$ and coherent for $\epsilon = 2.6$. There are more windows (not shown in figure) of chimera and multi-chimera beyond $\epsilon = 2.6$ and before the coupled system settle down to stable coherent state for high value of $\epsilon$.\\
We have checked with calculation that such pattern begin to show for both types of dynamics ($b=0.1998$ \& $b=0.18$) with linear coupling when $r = 0.05$. Pattern also start emerging for stable chaotic dynamics($b= 0.18$) with non-linear coupling with $r = 0.05$. $r = 0.03$ can be considered as nearly local. Therefore chimera and multi-chimera patterns emerge in the case of non-linearly coupled quasi periodic oscillators in a nearly local coupling architecture.

\section{Application to Self Propelled Particles Dynamics}

Self propelled coupled dynamical systems, like active Brownian motors, motile cells, macroscopic animals and artificial self-propelled particles\cite{roman} have directed motion due to velocity-velocity correlation. In this context variables $x$, $y$ and $z$ in Thomas model represent three components of velocity of active particle. Coupling the $x$ component only gives preferential directed motion. Theoretical tools for the description of such dynamical systems are mainly based on stochastic dynamics due to inherent random force coming from fluctuation in the environment. Therefore it is the probability distribution of velocity  is studied in such theoretical description. Thomas model in Chaotic or nearly chaotic regime does capture the random motion of active particles like chaotic walk in the phase portrait for $b = 0.0$, but in a deterministic manner. Therefore spatial velocity distribution in this model can be compared with probability distribution of velocity in stochastic model.\\ 
Stochastic model with non-linear velocity dependent friction\cite{roman} for active Brownian particles shows double crater like velocity distribution in Cartesian plane. The crater like distribution occurs due to asymmetry between forward and backward motion. The ring like distribution is due to preferred speed value in any direction\cite{roman}. Our result also show similar trend. Simultaneous existence of chimera and coherent states(Figure(\ref{AA})) or crater and coherent states(Figure(\ref{AD})) correspond to crater and ring like velocity distribution in Cartesian plane of stochastic model for active Brownian particle motion with non-linear velocity dependent friction.\\
Vicsek model(VM)\citep{T_vicsek} with vectorial noise\cite{Guill} for active self propelled particles predicts discontinuous transition from complete incoherent to coherent pattern with the decrease of noise strength($\eta$) in the limit of large size of the system. $\eta = 1$ corresponds to complete disorderness whereas at some critical low value, ${\eta}_c$ the active particles sudenly start moving in coherent manner. This shows first order phase transition. Particles have random walk motion for $\eta = 1$.  Thomas model with $b = 0$ predics also similar result for such a real system. $\epsilon = 0$ in Thomas model corresponds to incoherent state when individual particles undergoes random walk like motion in phase space. While at a critical value for $\epsilon$, there is a discontinuous transition to coherent state, showing phase transition. We are able to achieve this result by increasing the size of the system from 100 to 200 oscillators in accordance with the result of VM. VM model with vectorial noise and with cohesion\cite{Guill} consider local interaction which predicts cluster formation. Our result for Thomas model with local interaction also shows existence of coherent state.  

\section{Conclusion}
Network of hundred Thomas oscillators coupled with linear and sinusoidal non-linear coupling scheme is studied with varying coupling radius. It is known that chimera states is achieved for non-local coupling with radius around 0.2 for linearly coupled chaotic oscillators. In the present case also we are able to achieve chimera states for certain range of intermediate coupling constant with non-local coupling of radius 0.2. The system shows clusters, chimera and multi-chimera as the coupling is increased unlike the reversed order observed in the case of other studied network of chaotic oscillators with linear coupling. Unlike with linear coupling, we observe chimera states even for almost local coupling with three nearest neighbor coupling. The system shows complete synchronized state for global coupling scheme as expected.\\
In the context of dynamics of coupled self propelled active agents, the probability distribution of particles' velocity resembles like the present observation of chimera states . This result establishes the validity of the present model to describe dynamics of interacting active agents(particles). The discontinuous jump from incoherent to coherent state in the case of $b=0$ shows first order phase transition and agrees with other statistical model for self propelled particles. We conclude that Thomas system with non-linear coupling is a suitable model to describe self propelled active agents, where as with linear coupling it does not completely describe such real systems.

\bibliography{Bibliography}

\end{document}